\documentstyle[prl,aps,epsf,twocolumn]{revtex}

\begin{document}
\title{Collision-assisted Zeeman cooling of neutral atoms}

\author{Gabriele Ferrari}
\address{Laboratoire Kastler Brossel,
D\'epartement de Physique de l'Ecole Normale Sup\'erieure,\\ 24 rue Lhomond, 75231 Paris
CEDEX 05, France}

\date{\today}
\maketitle

\begin{abstract}
We propose a new method to cool gaseous samples of neutral atoms. The gas is
confined in a non dissipative optical trap in the presence of an homogeneous
magnetic field. The method accumulates atoms in the $m_F=0$ Zeeman sub-level.
Cooling occurs via collisions that produce atoms in $m_F\neq 0$ states. Thanks
to the second order Zeeman effect kinetic energy is transformed into internal
energy and recycling of atoms is ensured by optical pumping. This method may
allow quantum degeneracy to be reached by purely optical means.
\end{abstract}

\vspace{5mm}

Since the first proposal of optical pumping \cite{Kastler50} many authors have
focused on different schemes to cool samples by means of light fields
\cite{Haensch75,Pritchard83}. Laser cooling has produced significant results
both for macroscopic and microscopic systems.

Spontaneous anti-Stokes scattering has been used to cool molecular gasses,
fluid solutions and solid states systems \cite{Djeu81,Clark96,Epstein95}.
However these cooling schemes have intrinsic limitations in the attainable
temperature due to the reduction of anti-Stokes scattering at low
temperatures.

In the microscopic domain laser cooling has been very efficient on gases of
ions and neutral atoms \cite{Varenna91}. Associated with evaporative cooling
\cite{Hess86,Masuhara88}, laser cooling was an essential step towards
Bose-Einstein condensation of weakly interacting gases
\cite{Anderson95,Varenna98}. For neutral atoms polarization gradient cooling
in optical molasses reaches temperatures on the order of 10 $T_{\rm r}$
($T_{\rm r}=\hbar^2 {\bf k}^2/mk_B$ is the single-photon recoil temperature,
{\bf k} is the photon wave-number). Sideband cooling in 3D has reached 1.5
$T_{\rm r}$ and the limitation by multiple photon scattering has been
identified \cite{Kerman2000,Boiron96}. 3D sub-recoil cooling \cite{Lawall95}
has reached $T_{\rm r}$/20 for free atoms but suffers from a loss of
efficiency for trapped atoms at high density \cite{Lee96}. In contrast
evaporative cooling takes advantage of elastic collisions and does not have
limitations due to light scattering. It only suffers from a significant loss
of atoms during evaporation.

In this letter we propose a new cooling mechanism that combines elastic collisions,
inelastic collisions and optical pumping to efficiently cool samples of alkali atoms with
no loss of atoms. In principle the method is able to reach sub-recoil temperatures and
hold promises to reach quantum degeneracy by purely optical means. The key points of the
method are the following: {\it (i)} the gas is stored in an optical far-off resonance trap
(FORT) \cite{Miller94} in a well defined Zeeman sub-level of the electronic ground state,
{\it (ii)} an uniform magnetic field is applied to the sample and its magnitude is chosen
to have the second order Zeeman energy on the order of $k_BT$, {\it (iii)} inelastic
collisions produce atoms in different Zeeman sub-levels of higher energy transforming
kinetic energy into internal energy, {\it (iv)} atoms are then optically pumped back to
the initial state by an additional laser beam with suitable polarization. Such a cycle
removes an energy of the order of the second order Zeeman energy and cooling occurs at a
rate set by the collision rate.

Now we apply the method to an alkali atom with nuclear spin 3/2 (such as
$^7$Li, $^{23}$Na or $^{87}$Rb). The energy shifts of the different Zeeman
sub-levels in a uniform magnetic field $B$ can be calculated using:
\begin{equation}
\xi=\frac{\mu_B B}{2 \hbar \omega_{\rm HF}}=\frac{\omega_L}{\omega_{\rm HF}}
\end{equation}
where $\mu_B$ is the Bohr magneton, $\hbar\omega_L$ is the linear Zeeman shift
between adjacent Zeeman sub-levels and $\hbar \omega_{\rm HF}$ is the
hyperfine splitting of the electronic ground state. The Zeeman corrections to
second order in $\xi$ read:
\begin{equation}\label{ZeemanCorr}
\delta_{F,m_F}(\xi)=(-1)^F[\hbar \omega_{\rm HF} m_F\xi +\hbar \omega_{\rm
HF}(4-m_F^2)\xi^2].
\end{equation}

Assume that the gas is polarized in the level $|F=1, m_F=0\rangle$ and
sufficiently cold that only s-wave collisions take place \cite{s-wave}. The
projection of the angular momentum  on the quantization axis is conserved,
hence a colliding pair of atoms either remains in the same internal state (A),
or changes the internal state to the pair $|F=1, m_F=-1\rangle ~+ ~|F=1,
m_F=+1\rangle$ (B). Those collision have already been observed in
\begin{figure}[b]
\vspace{-0cm}
\epsfxsize=4.8cm
\centerline{\epsfbox{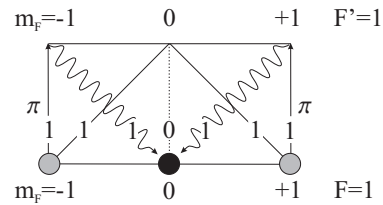}}           
\vspace{-0.4cm}
\caption{\label{LandeNegative} Scheme for the cooling in an hyperfine state with negative
Land\'e factor. Collisions between atoms in $|F=1, m_F=0\rangle$ (black round) produce
couples of atoms in $|F=1, m_F=\pm1\rangle$ of higher internal energy (grey). Energy
conservation in the collision is ensured by reduction of the kinetic energy of the
relative motion. Optical pumping on a $F=1\rightarrow F'=1$ transition with $\pi$
polarized light brings the atoms back to the initial state. In the sketch are indicated
the squares of the Clebsch-Gordan coefficients multiplied by 2.}
\end{figure}
\cite{Stenger98}. The total internal energy of the pairs A and B are equal to
first order in $\xi$ but differ to second order: $E_{\rm A}(\xi)=-8\xi^2 \hbar
\omega_{\rm HF}$, $E_{\rm B}(\xi)=-6\xi^2 \hbar \omega_{\rm HF}$. Since
$E_{\rm A}(\xi)<E_{\rm B}(\xi)$, couples of $|F=1, m_F=0\rangle$ atoms may
collide and change the internal state only when their energy in the center of
mass is greater than the energy threshold $\Delta=E_B(\xi)-E_A(\xi)$. This
endo-energetic collision (EC) transforms a fraction of kinetic energy into
internal energy. The cycling on EC's is insured by a $\pi$-polarized laser
resonant on a $F=1\rightarrow F'=1$ optical transition. $m_F=0$ atoms are not
coupled to the pumping light (see Fig. \ref{LandeNegative}) while $m_F=\pm1$
atoms once excited in $|F'=1, m_F=\pm1\rangle$ may decay with a 1/2 branching
ratio to A \cite{BranchingRatio}. The efficiency of the cooling process will
depend on the ratio of the removed energy and heating on each cycle: the
pumping process will heat the sample due to the recoil of the atom after the
absorption and spontaneous emission of the pumping photon. In fact two
different situation can be considered: the temperature is much larger than the
recoil energy $k_BT\gg E_{\rm r}$ ($E_{\rm r}=k_B T_{\rm r}/2$), or the two
are similar $k_BT\sim E_{\rm r}$ \cite{TemperatureLowerLimit}. In the first
case the heating associated with the pumping can be neglected and the cooling
is then equal to the rate of EC times the removed energy in each collision. In
the low energy domain the energy balance of the cooling will also include the
heating associated with the pumping process.

As stated  above, the cooling rate depends on the rate $\Gamma$ at which EC's
take place and the amount of the energy removed. To calculate $\Gamma$, we
consider the gas as a classical homogeneous gas and we determine the fraction
of collisions with energy in the center of mass $E_{\rm CM}$ greater than the
removed energy $\Delta$:
\begin{eqnarray}
\Gamma(\Delta,T)\propto  & \sigma & \int d{\bf p}_1\; d{\bf p}_2\;
n(1,T)n(2,T)\; \nonumber \\
 & \times & \Theta(E_{\rm CM}-\Delta)\;|{\bf v'}_1-{\bf v'}_2|,
\label{Gamm}
\end{eqnarray}
where $\sigma$ is the cross section for a collision changing the internal
state, 1 and 2 represent the two colliding atoms, $n(i,T)$ is the Boltzmann
factor for atom $i$ at temperature $T$, $\Theta$ is the step function equal to
1 if $E_{\rm CM}>\Delta$ and null otherwise, and ${\bf v'}_i$ is the velocity
of atom $i$ after the changing of the Zeeman sub-level. The factor $|{\bf
v'}_1-{\bf v'}_2|$ in equation \ref{Gamm} takes in account the reduction of
the density of final states when the energy in the center of mass changes
during the collision \cite{Landau}.

\begin{figure}[t]
\vspace{-1.6cm}
\begin{center}
\epsfxsize=7cm
\centerline{\epsfbox{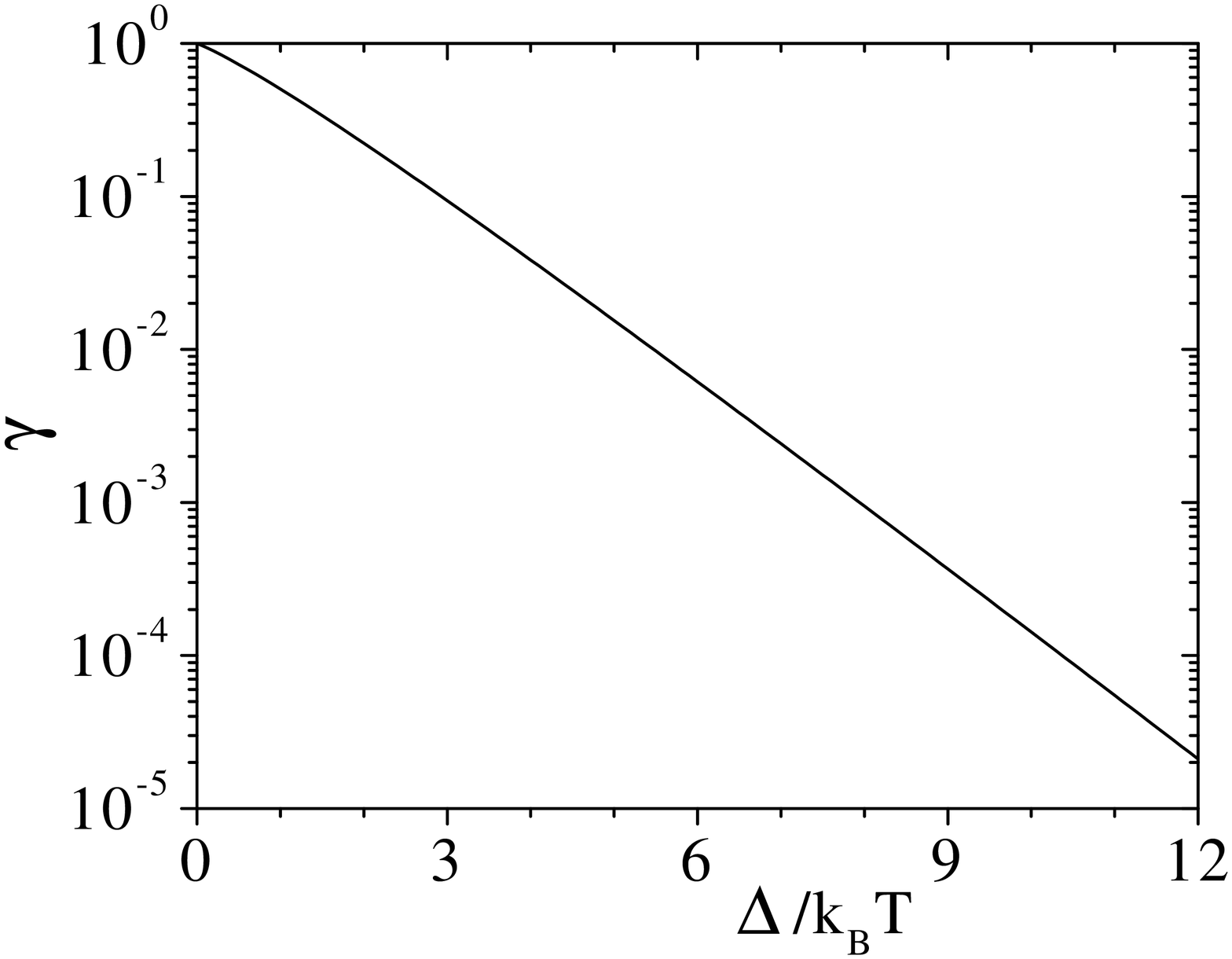}}
\end{center}
\vspace{-0.7cm} \caption{\label{CollisionalRate}
$\gamma(\Delta/k_BT)=\Gamma(\Delta,T)/\Gamma(0,T)$ is the normalized rate of
collisions with energy in the center of mass greater than the energy threshold
expressed in temperature units.}
\end{figure}

\begin{figure}[t]
\vspace{-1cm}
\begin{center}
\epsfxsize=7cm
\centerline{\epsfbox{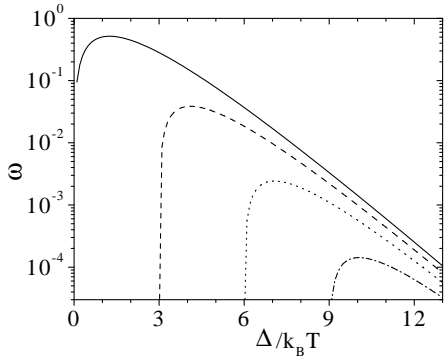}}
\end{center}
\vspace{-0.7cm} \caption{\label{CoolingRate} $\omega=(\Delta-E_{\rm
p})\gamma(\Delta/k_BT)/(k_BT) $ is the cooling rate in units of temperature
and collision rate at zero threshold. $E_{\rm p}=0$ (solid line), $E_{\rm
p}=3\,k_BT$ (dashes), $E_{\rm p}=6\,k_BT$ (dots), $E_{\rm p}=9\,k_BT$
(dash-dots).}
\end{figure}

Fig. \ref{CollisionalRate} shows $\gamma(\Delta/k_BT)$, {\it i.e.} the rate
$\Gamma(\Delta,T)$ normalized by the collisional rate at zero energy threshold
$\Gamma(0,T)$. If we define $E_{\rm p}$ as the energy gained during the
pumping process (typically few $E_{\rm r}$), in the limit where the pumping is
faster than $\Gamma(\Delta,T)$ the cooling rate $W$ reads:
\begin{equation}
W(\Delta,T,E_{\rm p})=\Gamma(\Delta,T)(\Delta-E_{\rm p}). \label{Cool}
\end{equation}
Those rates are plotted in Fig. \ref{CoolingRate} for different amounts of heating
associated with the pumping. One finds that for different values of $E_{\rm p}/k_BT$, the
cooling rate
is optimized when $\Delta\simeq k_B T + E_{\rm p}$. Then it is possible to
make an estimation of the dynamics of the cooling process by solving
\begin{eqnarray}
\frac{{\rm d}\,T}{{\rm d}\,t} & = & -\frac{(\Delta-E_{\rm p})}{6k_B}
\Gamma\left(\Delta,T\right) \label{Diff1}
\\
 & = & -\frac{T_{\rm in}}{6}\,\Gamma  (0,T_{\rm in})\,\gamma\left(1+\frac{E_{\rm
                              p}}{k_BT}\right), \label{eqation2}
\end{eqnarray}
where $T_{\rm in}$ is the initial temperature after the loading of the FORT.
The denominator in Eq. \ref{Diff1} takes into account the heat capacity of a
3D confined gas and the fact that the energy subtracted in EC is provided by
the two colliding atoms. In Eq. \ref{eqation2} we use the fact that if the
trapping potential is harmonic (like the bottom of a FORT) $\Gamma(0,T)\propto
T^{-1}$. The dynamics of the cooling behaves differently in the two
temperature regimes: if $E_{\rm p}/k_BT\ll 1$ the argument of $\gamma$ is
constant therefore the temperature decreases linearly. Once $E_{\rm p}/k_BT
\simeq 1$ the normalized collision rate $\gamma$ is no longer constant and
since it can be approximated by an exponential function (see figure
\ref{CollisionalRate}) the temperature decreases logarithmicaly with time.

\begin{figure}[t]
\vspace{-1cm}
\begin{center}
\epsfxsize=7cm
\centerline{\epsfbox{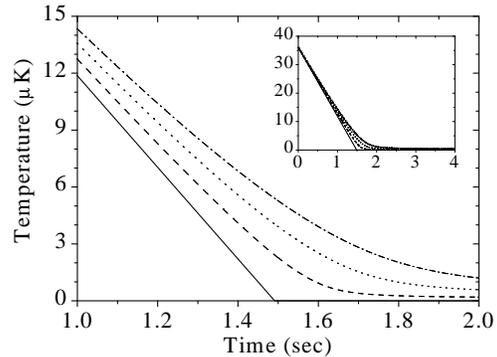}}
\end{center}
\vspace{-0.7cm} \caption{\label{Dynamic} Cooling dynamics of a gas of
$^{87}$Rb atoms initially at 36\,$\mu$K and 2\,10$^{11}$\,cm$^{-3}$ peak
density for different pumping energies $E_{\rm p}$. The gas is confined in an
harmonic potential and the cross section for EC's is supposed to be on the
same order as that for elastic collisions.  $E_{\rm p}=0$ (solid line),
$E_{\rm p}=6\, E_{\rm r}$ (dashes), $E_{\rm p}=12\, E_{\rm r}$ (dots), $E_{\rm
p}=18\, E_{\rm r}$ (dash-dots). The inset shows the same picture on a
different time scale.}
\end{figure}

As an example we consider a gas of $^{87}$Rb at 36\,$\mu$K and 2\,$10^{11}$\,cm$^{-3}$
peak density (typical values attainable after charging a FORT from a magneto-optical trap
\cite{Varenna91}). If we suppose that the collisional cross section for collisions
changing the Zeeman sub-level is of the same order as those for elastic collisions
\cite{Myatt97} we find $\Gamma(0,T_{\rm in})\sim4$\,s$^{-1}$. In Fig. \ref{Dynamic} are
plotted the integration of equation \ref{eqation2} for values of $E_{\rm p}$ ranging from
0 to 18\,$E_{\rm r}$ ($E_{\rm r}$=180\,nK for $^{87}$Rb). For this choice of parameters
the temperature drops almost linearly during the first two seconds. After the first 2
seconds typically we find $T\simeq E_{\rm p}/4 k_B $ and after 3 seconds $T\simeq E_{\rm
p}/6 k_B$. At longer cooling times $T\propto[\log t]^{-1}$ and reaches $E_{\rm p}/12k_B$
after 50 seconds. The present simulation will not be valid in the regime of extremely low
temperature when the gas can no longer be treated as a classical gas and when the
radiation trapping effect becomes extremely large. Experimentally $\Delta$ should be swept
from a value corresponding to the initial temperature of the gas, to slightly more than
$E_{\rm p}$. To avoid any dependency of the optical potential on $m_F$, the trapping light
has to be linearly polarized and propagating along the quantization axis. Slight
imperfections in the polarization of the trapping light do not affect the physics
discussed so far since they do not change the energy of colliding pairs of atoms if the
projection of the angular momentum is conserved.

An important issue for the cooling is the control of the magnetic field ${\bf
B}$. As stated before the direction of ${\bf B}$ has to be well defined and
parallel to the polarization of the pumping light. Indeed any misalignment
results in a $\sigma$ component in the polarization of the pumper that
introduces further heating terms. Since the $\pi$ component affects only
$m_F=\pm1$ atoms and the $\sigma$ component affects the ensemble of the gas,
choosing a Rabi frequency on the order of the rate of EC's will reduce the
heating due to the imperfect polarization. The requirements on the intensity
of ${\bf B}$ are much less demanding. In order to cool $^{87}$Rb ($\omega_{\rm
HF}=2 \pi\times 6.8$\,GHz) in the lower hyperfine state $F=1$, the magnetic
field necessary to have $\Delta$ corresponding to 100\,$\mu$K is 100\,Gauss.
The intensity of ${\bf B}$ is then swept down to slightly more than $E_{\rm
p}$, $\Delta\sim 1\mu$K (a few times $E_{\rm r}$). Since $\Delta \propto {\bf
B}^2$, the magnetic field at the end of the sweep will be of the order of
10\,Gauss. It is worth noting that loss due to light-assisted collisions such
as fine-structure or hyperfine-structure changing collisions can be avoided by
choosing properly the pumping transition: pumping of $^{87}$Rb on the
$F=1\rightarrow F'=1$ transition of the D1 line has no allowed channel for
those collisions.

\begin{figure}[t]
\hspace{-0.18cm}

\epsfxsize=7cm
\vspace{0cm}\centerline{\epsfbox{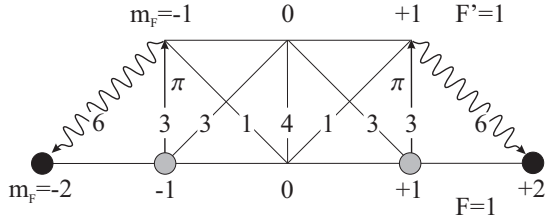}}

\vspace{-1.8cm} \caption{\label{LandePositive} Scheme for the cooling in an
hyperfine state with positive Land\'e factor. Atoms in $|F=2, m_F=\pm2\rangle$
(black round) collide producing couples of atoms in $|F=2, m_F=\pm1\rangle$ of
higher internal energy (grey) transforming part of the kinetic energy of the
relative motion into internal energy. Optical pumping on a $F=2\rightarrow
F'=1$ transition with $\pi$-polarized light brings the atoms back to the
initial states. In the sketch are indicated the squares of the Clebsch-Gordan
coefficients multiplied by 10.}
\end{figure}

Other cooling schemes are possible for the same atoms in the higher hyperfine
state (positive Land\'e factor $g_F$). Using Eq. \ref{ZeemanCorr} we find that
the  magnetic energy of the couple $|F=2, m_F=\pm1\rangle$ is $6\xi^2
\hbar\omega_{\rm HF}$ higher than that of couple $|F=2, m_F=\pm 2\rangle$.
Optical pumping on a $F=2\rightarrow F'=1$ transition with $\pi$-polarized light (see Fig.
\ref{LandePositive}) again permits cycling on EC's that lower $|m_F|$. A major obstacle to
this type of cooling is represented by inelastic collisions changing the hyperfine state.
Nevertheless it may remain
feasible for particular choices of atoms: $^{87}$Rb was proven to have the cross section
for spin-exchange collisions sufficiently small to make an unpolarized dilute cold gas
stable \cite{Myatt97}.

This cooling mechanism can also be extended to alkali atoms with nuclear spin different
from 3/2, in the highest hyperfine state. As in the previous case the atoms are polarized
in the extreme Zeeman levels $m_F=\pm F$ and EC's produce atoms in lower Zeeman levels.
The pumping is done by a $\pi$-polarized laser resonant on a $F\rightarrow F'=F-1$
transition. Other generalizations could consider mixtures of different atoms with
different $g_F$ factors. The cooling would then take advantage of the first-order Zeeman
effect.

The proposed cooling mechanism can be applied to produce Bose-Einstein
condensates \cite{Anderson95} with purely optical means. Since collisions
between atoms in different Zeeman sub-levels do not give rise to trap losses,
one can choose pumping rates arbitrarily low in order to fulfill the {\it
festina lente} scenario \cite{Cirac96} where the fluorescence rate is much
smaller than the oscillation frequency of the atom in the trapping potential.
The lower limit in the attainable temperature with this technique is
presumably due to the reabsorption of scattered photons during the pumping
phase. One possibility to partially avoid this effect is to reduce the
dimensionality of the trapping potential in order to have an elongated
cigar-shaped cloud \cite{Castin98}.

In conclusion we have presented a new, simple and spin-polarizing cooling
scheme that combines optical pumping, elastic collision and second order
Zeeman effect. The combination of optical pumping and endo-energetic
collisions permits to avoid the intrinsic limitation of laser cooling and
evaporative cooling: the finiteness of the exchangeable momentum between atoms
and photons, and the loss of atoms respectively. In presence of an homogeneous
magnetic field, collisions allow a defined amount of kinetic energy to be
transformed into internal energy (by changing the Zeeman state of the atoms)
and optical pumping ensures the cycling on the process.

Finally it is interesting to note that the proposed cooling method, polarizing the atoms
at will in the extremal Zeeman sub-level ($m_F=\pm F$) or $m_F=0$, is well adapted to
operation in atomic fountains. $^{87}$Rb, which seems to be a valid candidate as a future
frequency standard \cite{Bize99}, fits all the requirements of the presented mechanism.

I acknowledge fruitful discussions with  Iacopo Carusotto, Yvan Castin, Claude
Cohen-Tannoudji, Kristan Corwin, Jean Dalibard, Florian Schreck,  Christophe
Salomon and Guglielmo Tino. Laboratoire Kastler Brossel is an {\it Unit\'e de
recherche de l'Ecole Normale Sup\'erieure et de l'Universit\'e Pierre et Marie
Curie, associ\'ee au CNRS}. This work was partially supported by CNRS,
Coll\`{e}ge de France, DRET, DRED and EC (TMR network ERB FMRX-CT96-0002).


\begin{thebibliography}{99}
\bibitem{s-wave}
This temperature range is typically attained with laser cooling of alkali
atoms in magneto-optical traps.
\bibitem{BranchingRatio}
In fact the branching ratio towards the initial state is less since atoms may
relax towards the other hyperfine state with a branching ratio depending on
weather the optical repumping is made on the D1 or D2 line. Another repumping
beam should prevent optical pumping towards the wrong hyperfine state.
\bibitem{TemperatureLowerLimit}
We do not consider the case $k_B T \ll E_{\rm rec}$ since for realistic
experimental values in this case the gas would be Bose-condensed.
\end{thebibliography}
\end{document}